\documentstyle[12pt]{article}
\newcommand{\Pc}{Poincar\'{e} }
\newcommand{\2}{2+1}
\title{\nopagebreak
\vspace*{1.4in}
\large \bf Reduction of the Two Body Problem \\in \\ $N=2$ Chern
Simons supergravity}
\author{S. Kim and F. Mansouri \\  \it \small \it Physics
Department, University of
Cincinnati, Cincinnati, OH 45221}
\date{}

\begin{document}
\maketitle

\begin{abstract}
By an extension of the methods used for the reduction of the two
body problem in \2 dimensional gravity, we show that the two body
problem in N=2 Chern Simons supergravity can be reduced exactly to
an equivalent one body formalism.
We give exact expressions for the invariants of the reduced one
body problem.
\end{abstract}
\baselineskip=14pt
\bigskip
\par

In a recent work [1,2], it was shown that the two body problem for
sources of any spin in \2 dimensional gravity can be solved exactly
by reducing it to an equivalent one body problem. The work was
motivated by an earlier work of 't Hooft [3] who was the first to
recognize the possibility that the two boby problem can be solved
exactly.  To carry out the reduction, essential use was made of the
Chern Simons gauge theory of the \Pc group,the study of which was
initiated by Witten [4,5]. This formulation, aside from its
relevance to quantization, has the advantage of making the
topological properties of the theory manifest. These properties
were used to obtain, among other things, the mass and spin
associated with equivalent one body problem and the exact
expression for the two particle scattering amplitude [2,6].

The main object of this letter is to show that the two body problem
in $N=2$ supergravity theory in \2 dimensions is also
exactly solvable. Again we make use of the fact that supergravity
theories in \2 dimensions can also be formulated as Chern Simons
gauge theories of appropriate supersymmetric gauge groups [4,5].
Some features of the $N=2$ theory has been discussed in the
literature. The pure $N=2$ theory has been discussed by Dayi [7].
Its coupling to matter, the corresponding Wilson loop
observables,and application to superparticle scattering in the test
particle approximation was considered by Kohler,et al.[8,9]. To
establish our notation, we start with a brief summary of the
properties of the pure Chern Simons theory as given in references
[1] and [2].

The $N=2$ \Pc superalgebra in \2 dimensions can be written as
\begin{eqnarray}
& &[ J^{a},J^{b}] =-i\epsilon^{abc}J_{c} \hspace{.35in} ;
\hspace{.31in} [ P^{a},P^{b}]=0 \nonumber  \\
& &[ J^{a},P^{b}] =-i\epsilon^{abc}P_{c} \hspace{.31in} ;
\hspace{.31in} [ P^{a},Q_{\alpha}] =0  \nonumber \\
& &[ J^{a},Q_{\alpha}] =-(\sigma^{a})_{\alpha}^{\;\beta}Q_{\beta}
\hspace{.16in} ; \hspace{.29in} [ P^{a},Q'_{\alpha}] =0  \\
& &[ J^{a},Q'_{\alpha}] =-(\sigma^{a})_{\alpha}^{\;\beta}Q'_{\beta}
\hspace{.16in} ; \hspace{.3in} \{ Q_{\alpha},Q_{\beta}\} =0
\nonumber \\
& &\{ Q_{\alpha},Q'_{\beta}\} =-\sigma^{a}_{\;\alpha\beta}P^{a}
\hspace{.18in} ;
\hspace{.3in} \{ Q'_{\alpha},Q'_{\beta}\} =0 \nonumber \\
& & a=0,1,2 \hspace{.96in} ; \hspace{.34in} \alpha=1,2 \nonumber
\end{eqnarray}

The two component spinor charges $Q_{\alpha}$ and $Q'_{\alpha}$ are
raised and lowered by the antisymmetric metric
$\epsilon^{\alpha\beta}$ with $\epsilon^{12}=-\epsilon_{12}=1$. the
$SO(1,2)$ matrices $\sigma^{a}$ satisfy the Clifford algebra
\begin{equation}
\{\sigma^{a},\sigma^{b}\}=\frac{1}{2}\eta^{ab}
\end{equation}
where $\eta^{ab}$ is the Minkowski metric with signature $(+,-,-)$.
We also have
\begin{equation}
\sigma^{a}_{\;\alpha\beta}=(\sigma^{a})_{\alpha}^{\;\gamma}
\epsilon_{\gamma\beta}\end{equation}
A typical Chern Simons action in \2 dimensions is given by
\begin{equation}
I_{cs}=\int_{M} \gamma_{BC}A^{B}\wedge
(dA^{C}+\frac{1}{3}f^{C}_{\;DE}A^{D}\wedge A^{E})
\end{equation}
where $A^{B}$ are the components of the Lie superalgebra valued
connection
\begin{equation} A=A^{B}G_{B} \;\;\;\;  ;  \;\;\;\;
A^{B}=A^{B}_{\mu}dx^{\mu} \end{equation}
The quantity $\gamma$ is a suitable non-degenerate metric on the
superalgebra. For the \Pc superalgebra, its non-zero components are
given by [5]
\begin{equation}
<J^{a},P^{b}>=<P^{a},J^{b}>=\eta^{ab} \;\; ; \;\;
<Q^{\alpha},Q'^{\beta}>=<Q'^{\alpha},Q^{\beta}>=
\epsilon^{\alpha\beta}
\end{equation}
For this algebra, the connection can be written as
\begin{equation}
A_{\mu}=e_{\mu}^{\;a}P_{a}+\omega_{\mu}^{\;a}J_{a}+\chi_{\mu}^{\;
\alpha}Q_{\alpha}+\xi_{\mu}^{\;\alpha}Q'_{\alpha}
\end{equation}
Then with the covariant derivative
\begin{equation}
D_{\mu}=\partial_{\mu}+iA_{\mu}
\end{equation}
The components of the field strength tensor are given by
\begin{eqnarray}
F_{\mu\nu} & = & -i[D_{\mu},D_{\nu}] \nonumber\\
          &  = &
P_{a}[\partial_{\mu}e_{\nu}^{\;a}-\partial_{\nu}e_{\mu}^{\;a}
+\epsilon^{a}_{\;bc}(e_{\mu}^{\;b}\omega_{\nu}^{\;c}
+e_{\nu}^{\;c}\omega_{\mu}^{\;b}+i\sigma_{\alpha\beta}(\chi_{\mu}
^{\;\alpha}\xi_{\nu}^{\;\beta}+\xi_{\mu}^{\;\alpha}\chi_{\nu}^{\;
\beta})] \nonumber
\\
& &
+J_{a}[\partial_{\mu}\omega_{\nu}^{\;a}-\partial_{\nu}\omega_{\nu
}^{\;a}+\epsilon^{a}_{\;bc}\omega_{\mu}^{\;b}\omega_{\nu}^{\;c}]
\\
& &
+Q_{\alpha}[\partial_{\mu}\chi_{\nu}^{\;\alpha}-\partial_{\nu}
\chi_{\mu}^{\;\alpha}-i(\sigma_{a})_{\beta}^{\;\alpha}(\omega_
{\mu}^
{\;a}\chi_{\nu}^{\;\beta}-\chi_{\mu}^{\;\beta}\omega_{\nu}^{\;a})]
\nonumber \\
& &
+Q'_{\alpha}[\partial_{\mu}\xi_{\nu}^{\;\alpha}-\partial_{\nu}\xi
_{\mu}^{\;\alpha}-i(\sigma_{a})_{\beta}^{\;\alpha}(\omega_{\mu}^{
\;a}\xi_{\nu}^{\;\beta}-\xi_{\mu}^{\;\beta}\omega_{\nu}^{\;a})]
\nonumber
\end{eqnarray}
With these preliminaries,the Chern Simons action for the super \Pc
group can be written as
\begin{eqnarray}
I_{cs} &=& \frac{1}{2}\int_{M}\{ \eta_{bc}[e^{b}\wedge
(2d\omega^{c}+\epsilon^{c}_{\;da}\omega^{d}\wedge\omega^{a}]
\nonumber \\
& &
-\epsilon_{\alpha\beta}[\chi^{\alpha}\wedge(d-i\sigma_{a}\omega^{
a})\psi^{\beta}+\psi^{\alpha}\wedge(d-i\sigma_{a}\omega^{a})\chi^
{\beta}]\}
\end{eqnarray}
This action is invariant under the local gauge transformations
\begin{equation}
\delta A_{\mu}=\partial_{\mu}u+i\left[ A_{\mu},u \right]
\end{equation}
where
\begin{equation}
u=\rho^{a}P_{a}+\tau^{a}J_{a}+\nu^{\alpha}Q_{\alpha}+\nu'^{\alpha
}Q'_{\alpha}
\end{equation}

To set up a canonical formalism, let the manifold $M$ have the
topology $R\times \Sigma$, where $R$ is the real line representing
$x^{0}$ and $\Sigma$ is a two dimensional manifold parametrized by
$\{ x^{i}\}$ ,$i=1,2$. Then,up to total derivatives,the Chern
Simons action (10) reduces to [9]
\begin{eqnarray}
I_{cs} & = & \int dx^{0}\int_{\Sigma}\{
-\epsilon^{ij}e^{a}_{i}\partial_{x^{0}}\omega_{aj}
+\epsilon^{ij}\epsilon_{\alpha\beta}\chi^{\alpha}_{i}\partial_{x^
{0}}\xi^{\beta}_{j}\\
& & -\eta_{ab}(e^{a}_{0}F^{b}[\omega]+\omega^{a}_{0}F^{b}[e])
+\epsilon_{\alpha\beta}(\chi^{\alpha}_{0}F^{\beta}[\xi]+\xi^
{\alpha}_{0}F^{\beta}[\chi])\} \nonumber
\end{eqnarray}
where
\begin{equation}
F^{B}[A]=\epsilon^{ij}F^{B}_{\;ij}[A] \;\; ; \;\;
B=\left(a,\alpha\right)
\end{equation}
Analogously to the Chern Simons action for the \Pc group, the
structure of the action (13) indicates that the constraints of the
theory are
\begin{equation}
F^{a}[\omega]=F^{a}[e]=F^{\alpha}[\xi]=F^{\alpha}[\chi]=0
\end{equation}
Thus, the theory is locally trivial, as expected from any pure
Chern Simons theory.

To couple (super)sources to this Chern Simons theory, we proceed in
a manner similar to the way sources were coupled to the \Pc Chern
Simons theory [5,9,1,2]. It will be recalled that in the latter
case a source(particle) was taken to be an irreducible
representation of the \Pc group in \2 dimensions[1,2]. Its mass and
spin were identified as eigenvalues of the Casimir invariants of
the corresponding \Pc state. In generalizing this to the
supersymmetric case, we take a superparticle(supersource) to be an
irreducible representation of the super \Pc group. From this point
of view, a superparticle is an irreducible supermultiplet
consisting of several \Pc states related to each other by the
action of the supersymmetric generators. For N=2 super \Pc group,
the lowest dimensional supermultiplet consists of four \Pc states.
In this respect the supermultiplets in $2+1$ and $3+1$ dimensions
are similar. We note, however, that in contrast to the $3+1$
dimensional states, the spin of a \Pc state in \2 dimensions is not
limited to integer and half integer values. As a result, the spins
of states within a supermultiplet are not necessarily integer and
half integers. Of course, their spins must still differ from one
another by multiples of one-half unit. Here for definiteness we may
take the supermultiplet to be the N=2 vector supermultiplet
consisting of a spin zero, two spin one-half, and one spin one
states. But the action that we write down below for the coupling of
an irreducible N=2 supermultiplet to super Chern Simon thoery will
be invariant under super \Pc gauge transformations regardless of
the spin content of the supermultiplet.

To couple a supersymmetric source to the N=2 Chern Simons theory,
we recall [1,2] that a \Pc state is characterized by its monentum
$p^{a}=(p^{0},\vec{p})$ and its total(Lorentz) angular momentum
\begin{equation}
j^{a}=\epsilon^{a}_{\;bc}q^{b}p^{c}+s^{a}
\end{equation}
where $q^{a}$ are phase space coordinates canonically conjugate to
$p^{a}$, and $s^{a}$ determine the intrinsic spin. From these we
can construct the Casimir invariants $p^{2}=m^{2}$ and $W^{2}$ of
the \Pc group, where
\begin{equation}
W=p\cdot j=p\cdot s \; \; ; \; \; W^{2}=m^{2}s^{2}
\end{equation}
To realize the super \Pc algebra, we extend the phase space
variables $p^{a}$ and $q^{a}$ to their supersymmetric form:
\begin{equation}
p^{a}\longrightarrow \left( p^{a},p^{\alpha}\right) \;\; ; \;\;
q^{a}\longrightarrow \left( q^{a},q^{\alpha}\right)
\end{equation}
 In terms of these variables, the generators of the super \Pc
algebra take the form
\begin{eqnarray}
P_{a}&=& i\partial_{a}  \;\;\; ; \;\;\;
Q_{\alpha}=-\partial_{\alpha} \;\;\; ; \;\;\;
Q'_{\alpha}=i(\sigma^{a})_{\alpha\beta}q^{\beta}\partial_{a}
\nonumber\\
J_{a}&=& \epsilon_{abc}q^{b}p^{c}+(\sigma_{a})_{\alpha}^{\;
\beta}q^{\alpha}\partial_{\beta}+s^{a}
\end{eqnarray}
Then the source action can be written as
\begin{eqnarray}
I_{s}=\int_{C}d\tau\{
p_{a}\partial_{\tau}q^{a}-\epsilon_{\alpha\beta}p^{\alpha}
\partial_{\tau}q^{\beta}-e^{a}p_{a}-\omega^{a}j_{a}+i\epsilon_
{\alpha\beta}\chi^{\alpha}p^{\beta} \nonumber \\
-(\sigma \cdot
p)_{\alpha\beta}\xi^{\alpha}q^{\beta}+\lambda_{1}(p^{2}-m^{2})
+\lambda_{2}(C_{-}-ms)\}
\end{eqnarray}
where $\tau$ is an invariant parameter along the trajectory $C$.
This action is similar to the source action given in reference [9],
but it differs from the latter in the definition of $j_{a}$ and the
choice of the constraint multiplying $\lambda_{2}$. These features
turn out to be crucial in relating the second Casimir invariant,
$C_{-}$, of the superalgebra to the spin content of a
supermutiplet [10]. For more than one source, one can add an action
of this type for each source.

Under the gauge transformation (12), various quantities in the
action transform as follow :
\begin{eqnarray}
\delta e_{\mu}^{a} & = &
\partial_{\mu}\rho^{a}+\epsilon^{a}_{\;bc}(e_{\mu}^{b}\tau^{c}
+\omega_{\mu}^{b}\rho^{c})+i(\sigma^{a})_{\alpha\beta}(\chi_{\mu}
^{\alpha}\nu'^{\beta}+\xi_{\mu}^{\alpha}\nu^{\beta}) \nonumber \\
\delta \omega_{\mu}^{a} & = &
\partial_{\mu}\tau^{a}+\epsilon^{a}_{\;bc}\omega_{\mu}^{b}\tau^{c}
\nonumber \\
\delta \chi_{\mu}^{\alpha} & = &
\partial_{\mu}\nu^{\alpha}+i(\chi_{\mu}^{\beta}\tau^{a}-\omega_
{\mu}^{a}\nu^{\beta})(\sigma_{a})_{\beta}^{\; \alpha} \nonumber \\
\delta \xi_{\mu}^{\alpha} & = &
\partial_{\mu}\nu'^{\alpha}+i(\xi_{\mu}^{\beta}\tau-\omega_{\mu}^
{a}\nu'_{\beta})(\sigma_{a})_{\beta}^{\; \alpha}  \\
\delta q^{a} & = &
-\rho^{a}-\epsilon^{a}_{\;bc}\tau^{c}q^{b}-(\sigma^{a})_
{\alpha\beta}\nu'^{\alpha}q^{\beta} \nonumber \\
\delta q^{\alpha} & = & -i\nu^{\alpha}+i(\sigma \cdot
\tau)_{\beta}^{\; \alpha}q^{\alpha}  \nonumber \\
\delta p^{a} &  = & -\epsilon^{a}_{\;bc}\tau^{c}p^{b} \nonumber \\
\delta p^{\alpha} & = & -(\sigma \cdot p)_{\beta}^{\;
\alpha}\nu'^{\beta} + i(\sigma \cdot \tau)_{\beta}^{\;
\alpha}p^{\beta} \nonumber
\end{eqnarray}
It is easy to verify that the combined action $I=I_{cs}+I_{s}$ is
invariant under these infinitesimal gauge transformations.

Next, we turn to the reduction of the two body problem. We use this
terminology in the same sense that, e.g.\ , the familiar two body
central force problem can be reduced to an equivalent one body
problem. The main difference is that to maintain gauge invariance,
in the present case the reduction would have to be carried out in
terms of Wilson loops. The Wilson loop for the connection given by
Eq.\ (7) in the representation $R$ of the algebra is given by
\begin{equation}
W_{R}(C)=Str_{R}P\exp[i\oint_{C}A]
\end{equation}
where {\em Str\/} stands for supertrace and P for path ordering. We
have identified our superparticles with irreducible representations
of the super \Pc group, indicating that the Casimir invariants of
the super \Pc state are the observables of the superparticle. Since
all the gauge invariant observables of a Chern Simons theory are
Wilson loops,these Casimir invariants are expressible in terms of
Wilson loops. This is in particular true for the equivalent one
body state which is also a super \Pc state. Just as the one body
\Pc state was endowed with momenta $\Pi^{a}=(\Pi^{0},\vec{\Pi})$
and (Lorentz)angular momenta $\Psi^{a}=(\Psi^{0},\vec{\Psi})$
[1,2], we take the one body super \Pc state to be endowed with
charges
\begin{equation}
\Pi^{A}=(\Pi^{a},\Pi^{\alpha}) \;\;and \;\;
\Psi^{A}=(\Psi^{a},\Psi^{\alpha})
\end{equation}
The corresponding Casimir invariants are
\begin{equation}
C_{+}=\Pi^{a}\Pi_{a}\equiv H^{2} \; \; ; \; \;
C_{-}=\Pi^{a}\Psi_{a}+\epsilon^{\alpha\beta}\Pi_{\alpha}
\Psi_{\beta}
\end{equation}
where $H$ is the mass(Hamiltonian) of the supermultiplet. Being
gauge invariant observables of the unbroken supersymmetric gauge
theory, they can be expressed in terms of the Wilson loops of the
two body system.

In evaluating the invariants $C_{+}$ and $C_{-}$ of the super-\Pc
state representing the equivalent one body system in terms of a
Wilson loop of actual two body system, it turns out to be
technically more convenient to evaluate the corresponding Wilson
loop, $\hat{W}_{R}(C)$, for the super anti-de Sitter algebra
$OSp(1|2;R)\times OSp(1|2;R)$ and then obtain the super \Pc limit
by a group contraction. Let $X^{A}=(X^{a},S^{X}_{\alpha})$ and
$Y^{A}=(Y^{a},S^{Y}_{\alpha})$ be the generators of the two
commuting $OSp(1|2;R)$ subalgebras of the super anti-de Sitter
algebra with following non-zero (anti)commutators:
\begin{eqnarray}
& &[ X^{a},X^{b}] =-i\epsilon^{abc}X_{c} \hspace{.54in} ;
\hspace{.34in} [ Y^{a},Y^{b}] =-i\epsilon^{abc}Y_{c} \nonumber \\
& &[ X^{a},S^{X}_{\alpha}]
=-(\sigma^{a})_{\alpha}^{\;\beta}S^{X}_{\beta} \hspace{.37in} ;
\hspace{.34in} [ Y^{a},S^{Y}_{\alpha}]
=-(\sigma^{a})_{\alpha}^{\;\beta}S^{Y}_{\beta} \\
& &\{ S^{X}_{\alpha},S^{X}_{\beta}\} =-(\sigma \cdot
X)_{\alpha\beta} \hspace{.27in} ;\hspace{.34in}
\{ S^{Y}_{\alpha},S^{Y}_{\beta}\} =-(\sigma \cdot Y)_{\alpha\beta}
\nonumber
\end{eqnarray}
Then,let
\begin{eqnarray}
& & \hat{J}_{a}=X_{a}+Y_{a} \hspace{.68in} ; \hspace{.3in}
\hat{P}_{a}=\lambda(X_{a}-Y_{a}) \\
& & \hat{Q}_{\alpha}=\sqrt{\lambda}(S^{X}_{\alpha}+S^{Y}_{\alpha})
\hspace{.2in}
; \hspace{.3in}
\hat{Q}'_{\alpha}=\sqrt{\lambda}(S^{X}_{\alpha}-S^{Y}_{\alpha})
\nonumber
\end{eqnarray}
It is easy to verify that these operators satisfy the super anti-de
Sitter algebra. The caret on top of these operators as well as the
source charges,etc.\ ,given
below indicate that they correspond to the super anti-de Sitter
algebra. In the limit $\lambda \longrightarrow 0$, the super \Pc
algebra is recovered.
In particular, the Casimir operators ${C}_{\pm}$ of the super \Pc
algebra can be recovered in this limit :

\begin{equation}
\hat{C}_{\pm}=\hat{C}^{X}\pm \hat{C}^{Y} \longrightarrow C_{\pm}
\end{equation}
where
\begin{eqnarray}
C_{+} &=& P\cdot P \\
C_{-} &=& P\cdot J + J\cdot P +
\epsilon^{\alpha\beta}Q_{\alpha}Q'_{\beta}
\end{eqnarray}

Next, we turn to the computation of $\hat{W}_{R}(C)$. The source
which represents the equivalent one body problem is endowed with
super anti-de Sitter
charges
\begin{equation}
\hat{\Pi}^{A}=(\hat{\Pi}^{a},\hat{\Pi}^{\alpha}) \; \; \; ; \; \;
\;
\hat{\Psi}^{A}=(\hat{\Psi}^{a},\hat{\Psi}^{\alpha})
\end{equation}
These charges are subject to the requirement that after group
contraction,
$\hat{\Pi}^{A}\longrightarrow \Pi^{A},\hat{\Psi}^{A}\longrightarrow
\Psi^{A}$
such that
\begin{eqnarray}
C_{+} &=& \eta^{ab} \Pi_{a}\Pi_{b}  \\
C_{-} &=&
\eta^{ab}\Pi_{a}\Psi_{b}+\epsilon^{\alpha\beta}\Pi_{\alpha}
\Psi_{\beta}
\end{eqnarray}
Then, the Wilson loop around the source with charges given by (30)
will take the form
\begin{eqnarray}
\hat{W}_{R}(C) &=&
Str_{R}P\exp[i(\hat{\Pi}^{a}\hat{J}_{a}+\hat{\Psi}^{a}\hat{P}_{a}+
\hat{\Pi}^{\alpha}\hat{Q}_{\alpha}+\hat{\Psi}^{\alpha}\hat{Q'}_
{\alpha}] \nonumber   \\
&=&
Str_{R}P\exp[i(\frac{\hat{Z}^{A}_{+}}{2}X_{A}+
\frac{\hat{Z}^{A}_{-}}{2}Y_{A})]
\end{eqnarray}
where
\begin{equation}
\hat{Z}^{A}_{\pm}=(\hat{Z}^{a}_{\pm},\hat{Z}^{\alpha}_{\pm})
=(\hat{\Pi}^{a}\pm \hat{\Psi}^{a},\hat{\Pi}^{\alpha}\pm
\hat{\Psi}^{\alpha})
\end{equation}
Just as in the trace computation of \Pc Chern Simons theory [1,2],
the supertrace in (32) can be computed by considering a matrix
representation of the superalgebra.
The result is
\begin{equation}
W_{R}(C_{0})=
(2\cos\frac{|Z_{+}|}{2}-1)(2\cos\frac{|Z_{-}|}{2}-1)
\end{equation}
where, after group contraction,
\begin{equation}
|Z_{\pm}|=[\Pi^{a}\Pi_{a}\pm 2(\Pi^{a}\Psi_{a}+
\epsilon^{\alpha\beta}\Pi_{\alpha}\Psi_{\beta})]^
{\frac{1}{2}}
\end{equation}
To express $|Z_{\pm}|$ in terms of the properties of the two actual
sources,
we must set $W_{R}(C_{0})$ equal to a Wilson loop,$W_{R}(C_{12})$,
enclosing these sources. The path $C_{12}$ can be chosen uniquely
by the requirement
that the equivalent one body problem correctly give the asymptotic
observables of the emerging space-time theory, as in the case of
the \Pc theory [1,2]. It turns out that $C_{12}$ is the simple loop
enclosing the two sources. Let the two sources have charges
$(p_{1}^{A},j_{1}^{A})$ and $(p_{2}^{A},j_{2}^{A})$,respectively.
Again we begin with the super anti-de Sitter Wilson loop,
$\hat{W}_{R}(C_{12})$, and obtain
$W_{R}(C_{12})$ by group contraction. We have
\begin{equation}
\hat{W}_{R}(C_{12})=StrP\prod_{k=1}^{2}\exp\left[ i(\hat{p}_{k}
\cdot \hat{J}+
\hat{j}_{k} \cdot
\hat{P}+\hat{p}^{\alpha}_{k}\hat{Q}_{\alpha}+\hat{j}^{\alpha}_{k}
\hat{Q'}_{\alpha})\right]
\end{equation}
Evaluating this expression and performing the subsequent group
contraction,
we obtain
\begin{equation}
W_{R}(C_{12})=W_{x}W_{y}
\end{equation}
where
\begin{eqnarray}
W_{x} &=& \{1
 +2(\cos\frac{|x_{1}|}{2}-1)+2(\cos\frac{|x_{2}|}{2}-1) \nonumber
\\
& & -2(|x_{1}||x_{2}|)^{-1}(x^{a}_{1}\cdot
x^{a}_{2}-\epsilon_{\alpha\beta}q^{\alpha}_{1}q^{\beta}_{2})
\sin\frac{|x_{1}|}{2}\sin\frac{|x_{2}|}{2} \nonumber \\
& &
+(|x_{1}||x_{2}|)^{-2}(\cos\frac{|x_{1}|}{2}-1)
(\cos\frac{|x_{2}|}{2}-1) \\
& & \times
[\;|x_{1}|^{2}|x_{2}|^{2}+4i\epsilon_{abc}x^{a}_{1}x^{b}_{2}
(\sigma)_{\alpha\beta}
q^{\alpha}_{1}q^{\beta}_{2}+(x^{a}_{1})^{2}(x^{a}_{2})^{2}
\nonumber \\
& & \;\;\;\;\;\;
-2x^{a}_{1}x^{a}_{2}\epsilon_{\alpha\beta}q^{\alpha}_{1}q^{\beta}
_{2}
+\epsilon_{\alpha\beta}\epsilon_{\gamma\delta}q^{\alpha}_{1}q^
{\gamma}_{2}q^{\beta}_{1}q^{\delta}_{2}\;] \} \nonumber
\end{eqnarray}
In this expression,with $k=1,2$, and $x^{A}=(x^{a},q^{\alpha})$, we
have
\begin{equation}
x^{a}_{k}=p^{a}_{k}+j^{a}_{k} \; \;\; ; \;\; \;
q^{\alpha}_{k}=p^{\alpha}_{k}+
j^{\alpha}_{k}
\end{equation}
The factor $W_{y}$ has exactly the same structure as $W_{x}$,
i.e., $x^{A}
\longrightarrow y^{A}=(y^{a},q'^{\alpha})$,where
\begin{equation}
y^{a}_{k}=p^{a}_{k}-j^{a}_{k} \; \;\; ; \;\; \;
q'^{\alpha}_{k}=p^{\alpha}_{k}-
j^{\alpha}_{k}
\end{equation}
Having evaluated the Wilson loops, we can now set
\begin{equation}
W_{R}(C_{0})=W_{R}(C_{12})
\end{equation}
to obtain the Casimir invariants $C_{\pm}$ of the supermultiplet
representing
the equivalent one body formalism. Since $C_{+}$ and $C_{-}$ are
independent,
consider first $C_{+}$ given by (24). To evaluate it, we set the
charges $j^{A}$ and $ p^{\alpha}$, which contribute to $C_{-}$ but
not to $C_{+}$, equal to zero. Then, from (35), (39), and (41), we
obtain
\begin{equation}
\cos\frac{H}{2}=\cos\frac{m_{1}}{2}\cos\frac{m_{2}}{2}-
\frac{p_{1}\cdot
p_{2}}{m_{1}m_{2}}\sin\frac{m_{1}}{2}\sin\frac{m_{2}}{2}
\end{equation}
Not surprisingly, this is exactly the same expression that was
obtained for
the mass of the equivalent one body state in the \Pc Chern Simons
theory[1,2].
This is as it should be because the quantity $H$ is also the common
mass of the \Pc states within our supermultiplet and could be
evaluated from the two body system independently of the
supersymmetry. Making use of this {\it a posteriori} knowledge, the
extraction of equation (43) from (35),
(39), and (41) may viewed as a test of the correctness of our
reduction formalism.

Having evaluated $H$ ,i.e., $C_{+}$, from equation (43), we can
substitute it into (35) and use (39) and (49) to obtain an exact,
albeit implicit expression for $C_{-}$.
A more detailed discussion of the properties of the supersymmetric
two body
system and its applications will be given elsewhere[10].

\bigskip
This work was supported in part by the Department of Energy under
the contract No.\ DOE-FG02-84ER40153.

\vspace{1in}
\noindent{\bf References}

\begin{enumerate}
\baselineskip=12pt

\item F. Mansouri and M.K. Falbo-Kenkel, {\it Mod. Phys. Lett.}
{\bf A8} (1993) 2503
\item F. Mansouri, {\it Proceedings of XVIIth Johns Hopkins
Workshopon on Current Problems in Particle Theory}, ed. Z. Horwath,
World Scientific, 1994
\item 10. G. 't Hooft, {\it Comm. Math. Phys.} {\bf 117} (1988) 685
\item A. Achucarro and P.K. Townsend, {\it Phys. Lett} {\bf B180}
(1986) 89
\item E. Witten, {\it Nucl. Phys.} {\bf B311} (1988) 46 and {\bf
B323} (1989) 113
\item M.K. Falbo-Kenkel and F. Mansouri, {\it Phys. Lett.} {\bf
B309} (1993) 28
\item O.F. Dayi, {\it Phys. Lett.} {\bf B234} (1990) 25
\item K. Koehler, F. Mansouri, C. Vaz, L. Witten, {\it Nucl. Phys.}
{\bf B341}
(1990) 167
\item K. Koehler, F. Mansouri, C. Vaz, L. Witten, {\it Nucl. Phys.}
{\bf B348}
(1990) 373
\item Sunme Kim and F. Mansouri, in preparation
\end{enumerate}
\end{document}